\documentclass[american,aps,pra,reprint, superscriptaddress, onecolumn]{revtex4-1}
\usepackage{amsmath,amscd,amsthm,amssymb}
\usepackage{mathrsfs}
\usepackage{graphicx}
\usepackage{epsf,epstopdf}
\usepackage[all]{xy}
\usepackage[unicode=true,pdfusetitle, bookmarks=true,bookmarksnumbered=false,bookmarksopen=false, breaklinks=false,pdfborder={0 0 0},backref=false,colorlinks=false] {hyperref}
\hypersetup{colorlinks=true,linkcolor=myurlcolor,citecolor=myurlcolor,urlcolor=myurlcolor}
\usepackage{graphics,epstopdf,graphicx, amsthm, amsmath, amssymb, times, braket, color, bm}
\usepackage{cleveref}
\usepackage{caption}
\usepackage{subcaption}
\usepackage{makecell}

\definecolor{myurlcolor}{rgb}{0,0,0.7}

\theoremstyle{plain}

\providecommand{\theoremname}{Theorem}
\newcommand*{\myproofname}{Proof}

\makeatother

\theoremstyle{definition}

\theoremstyle{remark}

\usepackage{amscd}

\newcommand{\beq}{\begin{equation}}
\newcommand{\eeq}{\end{equation}}
\newcommand{\ba}{\begin{array}}
\newcommand{\ea}{\end{array}}
\newcommand{\bea}{\begin{eqnarray}}
\newcommand{\eea}{\end{eqnarray}}

\begin{document}

\title{Matrix encoding method in variational quantum singular value decomposition}

\author{Alexander I. Zenchuk}
\email{zenchuk@itp.ac.ru}
\affiliation{Federal Research Center of Problems of Chemical Physics and Medicinal Chemistry RAS,
Chernogolovka, Moscow reg., 142432, Russia}
\author{Wentao Qi}
\email{qiwt5@mail2.sysu.edu.cn}
\affiliation{Institute of Quantum Computing and Computer Theory, School of
Computer Science and Engineering,
Sun Yat-sen University, Guangzhou 510006, China}
\author{Junde Wu}
\email{wjd@zju.edu.cn}
\affiliation{School of Mathematical Sciences, Zhejiang University, Hangzhou 310027, PR~China}

\begin{abstract}
We propose the variational quantum singular value decomposition based on  encoding the elements of the considered { $N\times N$} matrix  into the state of a quantum system of appropriate dimension. This method doesn't use the expansion of this matrix in terms of the unitary matrices. Controlled measurement is involved to avoid small success probability in ancilla measurement. The objective function for maximization algorithm can be obtained probabilistically via  measurement of the states of  { two}  one-qubit subsystems. The circuit requires $O(\log N)$ qubits for  realization of this algorithm  { whose depths is proportional to $ \log N/\varepsilon$,  where $\varepsilon$ is the  precision  required for calculation of singular values.}
\end{abstract}

\maketitle

\newpage

\section{Introduction}
Quantum algorithms { has become}  an attractive field of practical realization of quantum physics in everyday life.
Among other quantum  algorithms we concentrate on the family of algorithms manipulating { with}  matrices and thus representing the quantum analogues of classical counterparts. Quantum Fourier transform \cite{QFT1,QFT2,QFT3} and phase estimation \cite{QFT3,Wang} must be distinguished as most popular and well recognized quantum algorithms used as subroutines included in many algorithms  for data processing.

Set of algorithms reaches the goal using exclusively quantum approach.  Apart from Fourier transform and Quantum phase estimation quoted above we refer to
the algorithms for matrix manipulations (addition, multiplication,  Kronecker sum, tensor product, Hadamard product) based on the Trotterization method \cite{Trotter,Suzuki,T,AT} and the Baker-Champbell-Hausdorff   \cite{BC} approximation  for exponentiating matrices \cite{ZhaoL1}.  In \cite{QZKW_arxive2022}, the matrix operations (addition, multiplication, inner product of vectors) are realized via action of special unitary operations  on the matrix encoded into the  either mixed or pure superposition  state of some quantum system. Later, such unitary transformations where realized in terms of simple one- and two-qubit operations \cite{ ZQKW_arxive2023}.  The matrix-encoding approach was implemented in the quantum algorithms  for determinant calculation, matrix inversion and solving linear systems \cite{ZBQKW_2024}.

Another large family of algorithms  includes   algorithms combining  both quantum and classical subroutines. To such algorithms one can refer {  the original version of} the well known  Harrow-Hassidim-Lloyd (HHL) algorithm
for solving systems of linear algebraic equations \cite{HHL,HHL1,HHL2,HHL3,HHL4,HHL5,HHL6,HHL7}. {  In this algorithm the classical subroutine is required for inversion of eigenvalues $\lambda_j$ of the matrix $A$ in equation
$A{\boldsymbol x} ={\boldsymbol b}$. However, later the method of approximate calculation of  the inverse eigenvalues $1/\lambda_j$  was developed \cite{MRTC, NJ}. This method can be incorporated into HHL algorithm removing
necessity of classical operations.} Variational algorithms represent  widely acknowledged
class of hybrid algorithms for solving problems based on optimization methods. In particular, such algorithm was developed for the
 singular value decomposition (SVD) \cite{GSLW,BPGML,WSW,JKC}, where the loss (or objective) function at fixed optimization parameters was calculated by the quantum algorithm, while the iteration of parameters was performed using the classical gradient method. In Ref.\cite{WSW}, all simulations were implemented via Paddle Quantum \cite{Paddle1} on the PaddlePaddle Deep Learning Platform \cite{Paddle2,MWW}.

 We have to note that some principal issues on quantum algorithms for SVD are referred to Refs.\cite{KP,RSML,BLZ,KP2}.
 {  Thus, the principle of realization of the quantum gradient descent algorithm is described in \cite{KP}, but no explicit
representation for unitary transformation performing   iteration steps in this algorithm  is given, this problem requires  further study.
In \cite{RSML}, the problem of fixing the phase of the singular vector associated with the appropriate  singular value  is explored.
The data analysis involving the quantum SVD-based data representation is proposed in \cite{BLZ}.
The quantum singular value estimation algorithm
(i.e., estimation of the singular value associated with each singular vector) is described in \cite{KP,KP2}.}
However, the particular realizations of quantum algorithm are not discussed there. This fact motivates research on development of quantum  SVD algorithms which can be validated on near-term quantum processors.  The importance of the algorithms for SVD is determined by the wide applicability of SVD as a subroutine  in various algorithms   including some variants of matrix inversion \cite{MRTC, NJ}, solving systems of linear equations \cite{WZP}, quantum recommendation systems \cite{KP,ZDNarxive,BKParxive}.

 In our paper, we modify the algorithm developed in \cite{WSW,JKC} implementing the encoding the elements of the $N\times N$ matrix $A$  into the probability amplitudes of the  pure state of some quantum system with subsequent application of two parametrised unitary transformations  and matrix multiplications using the algorithm proposed in \cite{ ZQKW_arxive2023}. Implementing  this encoding we avoid representation of $A$ as a linear combination of unitary transformations utilised in  \cite{WSW,JKC}. In comparison with the above references, we reduce the number of measurements required to get the complete information for calculating the objective function. { In our case,  each run of the algorithm  is supplemented with $O(1)$ measurements of the uncillae states   (subsystems $K$ and $B$  below), while the number of measurements  in  \cite{WSW,JKC}  is $O(N)$, here $N$ is the number of diagonal elements in the  considered matrix. Of course, because of the probabilistic method of obtaining the objective function,  one has to perform series of runs of the algorithm.  }



The paper is organized as follows. In Sec.\ref{Section:SVD} we present the detailed description of  our version of the quantum part of the variational  SVD and briefly  discuss  the complete hybrid algorithm.
Conclusions are given in Sec.\ref{Section:Conclusions}.

\section{Singular value decomposition}
\label{Section:SVD}
\subsection{Preliminaries} 
\begin{figure}[ht]
\centerline{    \includegraphics[width=0.1\textwidth]{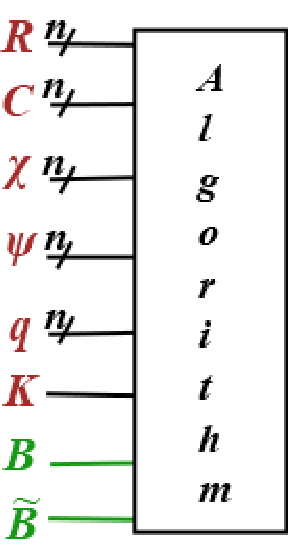}}
    \caption{{ Structure of subsystems required for performing the quantum part of variational  SVD. It includes five $n$-qubit subsystems for encoding the matrix $M$ (subsystems $R$, $C$), orthonormal eigenvectors  $\langle \psi_j|$ and $|\psi_j\rangle$   (subsystems $\chi$, $\psi$) and weights $q_i$ (subsystem $q$). In addition, the one-qubit subsystem $K$ serves as a controlling qubit in controlled operators of the algorithm and two one-qubit ancillae
    $B$ and $\tilde B$ serve  for the controlled measurement.}  }
\label{Fig:structure}
\end{figure}

We consider the singular value decomposition of an arbitrary square $N\times N$ matrix $M$ assuming $N=2^n$,
\begin{eqnarray}\label{SVD}
M=\hat U D \hat V^\dagger,
\end{eqnarray}
where $D={\mbox{diag}}(d_1,\dots,d_N)$ is the diagonal matrix of singular values (some of those values might be zero) and ${ \hat U}$ and ${ \hat V^\dagger}$ are unitary matrices.
To find the singular values (entries of the diagonal matrix $D$) we, first of all, introduce the
 following objective function \cite{WSW}:
\begin{eqnarray}\label{obj}
L(\alpha,\beta) = \sum_{j=0}^{N-1} q_j \times{\mbox{Re}} \Big(\langle \psi_j| U^T(\alpha) M U(\beta)|\psi_j\rangle \Big),\;\;U=\{b_{ij}:i,j=0,\dots,N-1\},
\end{eqnarray}
where $|\psi_j\rangle$, $j=0,\dots, N-1$, is {  a} set of orthogonal vectors, $\alpha=\{\alpha_0,\dots,\alpha_{nQ}\}$ and $\beta= \{\beta_0,\dots,\beta_{nQ}\}$ represent two sets of optimization parameters, $q_0>\dots>q_{N-1}$ are real  weights,  $Q$ is some integer associated with the subroutine used for preparing the unitary transformation $U$ in Sec.\ref{Section:W2W2}. We also note that the sum in (\ref{obj}) is over all $N$  singular values including possible zeros unlike Ref.\cite{WSW}, where the sum is truncated keeping only $T\le N$ largest singular values.
This truncation can be simply realized in our algorithm just equating to  zero all $q_j$ with $j>T$. The reason to take transposition $T$ of the matrix $U$ in (\ref{obj})  will be clarified letter, see eq.(\ref{AM}).
{  Here we emphasize that, although the objective function is a sum of $N$ terms, its expectation value  will be found by   measuring the states of  two qubis, see eqs.(\ref{tPsiOut}), (\ref{ppp}), unlike Refs. \cite{WSW,JKC}, where the  expectation value of  each term must be measured separately. }
We appeal to  the gradient method  to  find such parameters $\alpha^*$ and $\beta^*$ that maximize the objective function,
i.e.,
\begin{eqnarray}
\max_{\alpha,\beta} L(\alpha,\beta)  = L(\alpha^*,\beta^*).
\end{eqnarray}
Then
\begin{eqnarray}\label{UDV}
&&
d_j = \langle \psi_j| U^T(\alpha^*) M U(\beta^*)|\psi_j\rangle, \;\;j=0,
\dots,N-1,\\\nonumber
&&
\hat U^\dagger= U^T(\alpha^*),\;\;\hat V= U(\beta^*).
\end{eqnarray}
As for the set of orthogonal vectors $|\psi_j\rangle $, we take the vectors of computational basis $|j\rangle$, $j=0,\dots,N-1$, where  the integer $j$ is written in the binary form.  We introduce five $n$-qubit subsystems, see Fig.\ref{Fig:structure}:  the subsystems $R$ and $C$ serve to enumerate, respectively, the rows and column of $M$, the subsystems $\chi$ and $\psi$ are needed   to  operate with   $\langle \psi_j|$ and $|\psi_j\rangle$ in (\ref{obj}), these two subsystems  are also used  to organize the weighted  sum over $j$ in (\ref{obj}), and the subsystem $q$  encodes the normalized vector of weights,
\begin{eqnarray}\label{q}
|\varphi\rangle_q =\sum_{j=0}^{N-1} q_j|j\rangle_q, \;\; \sum_{j=0}^{N-1} q_j^2 =1.
\end{eqnarray}
In addition, the single qubit of the subsystem $K$ serves as a controlling qubit in  succeeding  controlled operations.  At the last steps of the algorithm we will introduce two one-qubit ancilae $B$ and $\tilde B$ to properly organize garbage removal and required measurements.

\noindent

We shall note that the proposed  method  can be also used to construct the eigensystem for the square positive semidefinite matrix  $R$ (for intance, for the density matrix), i.e., we can factorize $R$ as $R= U D U^\dagger$.
For this purpose, we  just have to involve the following relation between the matrices $U(\alpha)$ and
$U(\beta)$:
\begin{eqnarray}
\hat U^\dagger = U^T(\alpha)=\hat V^\dagger = U^\dagger(\beta) ,\;\; \Rightarrow  \;\;
  U(\alpha)= \bar U(\beta),
\end{eqnarray}
where the bar means complex conjugate.
This condition can be simply satisfied in the case when all the parameters $\alpha_j$ and $\beta_j$ are introduced via the $x$-, $y$- or $z$-rotation $R_{\theta j} =e^{-i \sigma^{(\theta)}\alpha_j/2 }$, $\theta=x,y,z$, $\sigma^{(\theta)}$ are the Pauli matrices. If $\theta=y$, then  $\alpha_j = \beta_j$. In the case
$\theta =x,z$, we take $\alpha_j=-\beta_j$.

\subsection{Quantum algorithm preparing objective function}
\label{Section:general}
First of all, we have to prepare the above mentioned matrix $M=\{m_{ij}: i,j=0,\dots,N-1\}$ for encoding into the state of a quantum system.
To this end we normalize  $M$  and make real the first diagonal element { assuming that this element does not equal to zero ($m_{00}\neq 0$),} i.e., replace  $M$ with the matrix $A=\{a_{ij}: i,j=0,\dots,N-1\}$:
\begin{eqnarray}\label{A}
A=\frac{e^{\displaystyle - i {\mbox{arg}}(m_{00})}}{\sqrt{ \sum_{i=0}^{N-1}  \sum_{j=0}^{N-1} |m_{ij}|^2}} M.
\end{eqnarray}
Now we can  encode the elements  of the matrix $A$ into the superposition state of $R$ and $C$ as follows:
\begin{eqnarray}\label{M}
&&
|A\rangle = \sum_{i=0}^{N-1}\sum_{j=0}^{N-1} a_{ij} |i\rangle_R |j\rangle_C ,\\\nonumber&&
 \sum_{i=0}^{N-1}\sum_{j=0}^{N-1} |a_{ij}|^2=1,
\end{eqnarray}
where the normalization is provided by  eq.(\ref{A}). In other words, we have quantum access to the matrix  $A$ \cite{BLZ}.
{ Here we shall  note that the  matrix encoding problem is a complicated task by itself  and, in general, the depth of the   algorithm encoding the $N\times N$ matrix  is at least $O(N^2)$.  The same holds for  encoding the state $|\varphi\rangle_q$ in (\ref{q}).  However, both this problems can be referred to the initial state preparation and will not be detailized in our paper.}
Subsystems $\chi$,  $\psi$ and $K$  are  in the ground state initially and the state of $q$ is defined in (\ref{q}), i.e., the initial state of the whole system { reads}:
\begin{eqnarray}\label{Phi0}
|\Phi_0\rangle=|A\rangle |0\rangle_\chi   |0\rangle_\psi
|\varphi\rangle_q  |0\rangle_K .
\end{eqnarray}
Hereafter in this paper the subscript means the subsystem to which the operator is applied.

\begin{figure}[ht]
\centerline{    \includegraphics[width=0.8\textwidth]{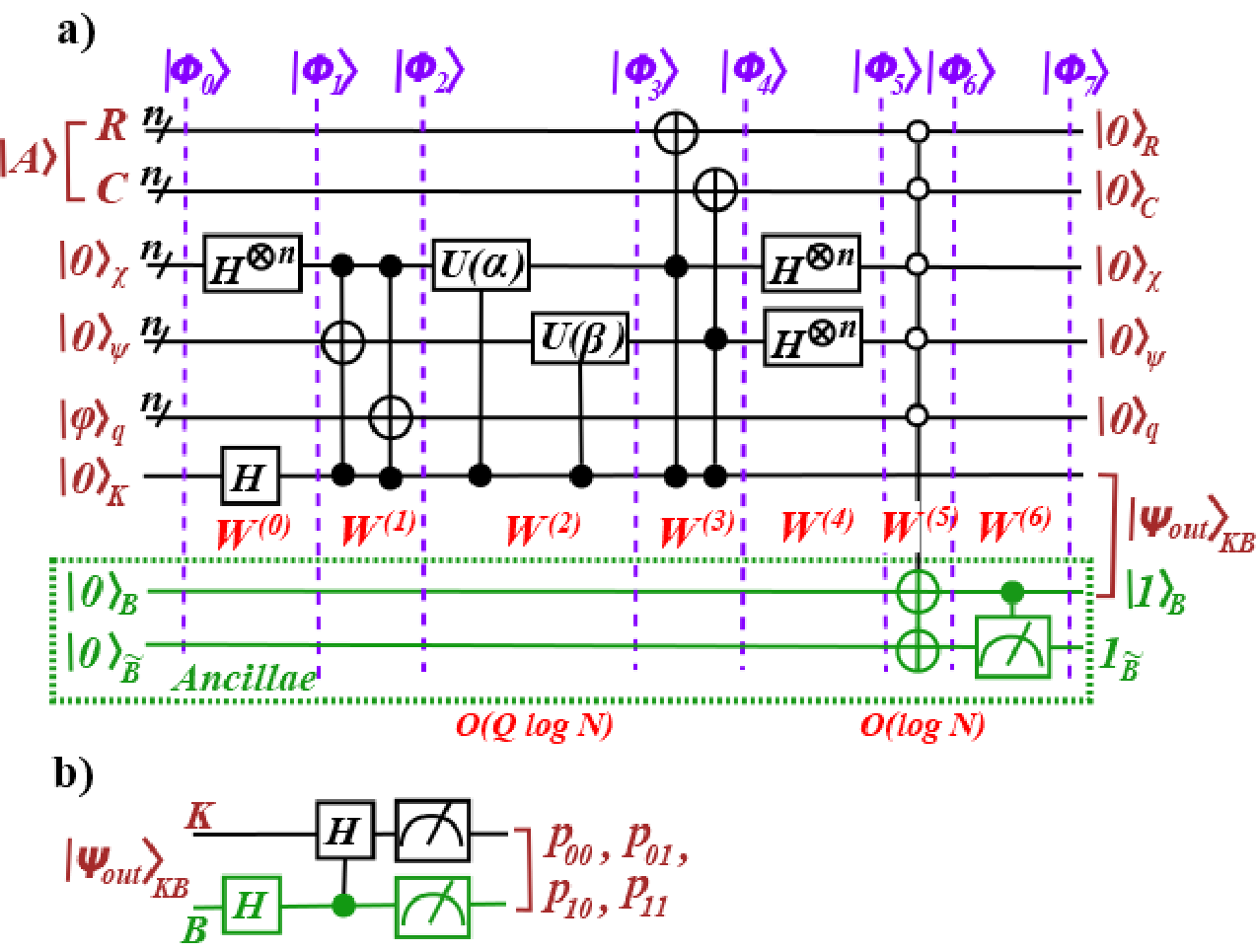}}
    \caption{Circuit for the quantum part of the variational SVD algorithm.  The depth of this circuit can be estimated as { $O(Q\log(N)/\varepsilon)$.} (a) The circuit for creating the state $|\Psi_{out}\rangle_K$ given in (\ref{Phi4}); the operators $W^{(j)}$, $j=0,\dots, 6$ are presented without subscripts  for brevity. (b) The operators applied to the state $|\Psi_{out}\rangle_{KB}$ to probabilistically  obtain the normalization $G$ and the objective function $L(\alpha,\beta)$. }
\label{Fig:SVD}
\end{figure}

Now we proceed to  {  description of} the quantum algorithm, which is also illustrated by  the circuit in Fig.\ref{Fig:SVD}.
As the first  step, we apply the Hadamard transformation to each  qubit of $\chi$ and $K$ (we denote this set of transformations as $
W^{(0)}_{\chi K}= H_\chi  H_K$):
\begin{eqnarray}\label{Phi1}
|\Phi_1\rangle = W^{(0)}_{\chi K} |\Phi_0\rangle =\frac{1}{2^{(n+1)/2}}
\sum_{k=0}^{N-1} |A\rangle |k\rangle_\chi   |0\rangle_\psi  |\varphi\rangle_q  ( |0\rangle_K+|1\rangle_K),
\end{eqnarray}
 thus creating  the  systems of orthonormal states $ |k\rangle_\chi$, $k=0,\dots, N-1$, and initializing the superposition state of the controlling qubit $K$.

Now we double the state  $|k\rangle_{\chi}$ creating the same state  $|k\rangle_{\psi}$ of the system $\psi$  and also multiply the obtained state by the weight $q_k$ resulting in  $q_k |k\rangle_{\chi} |k\rangle_{\psi} |0\rangle_{q}$, see eq.(\ref{Phi2}). In the last case we use the trick  proposed in \cite{ZBQKW_2024} for matrix product.  Both operations are
  controlled  by the  state of $K$, { i.e., they are applied only if $K$ is in the excited state
  $|1\rangle_K$.}
{  To arrange such control} we introduce the projectors
\begin{eqnarray}\label{PP}
P_{\chi_i K} = |1\rangle_{\chi_i} |1\rangle_K \,  _{\chi_i}\langle 1|_K\langle 1|,\;\; i=1,\dots, n,
\end{eqnarray}
and the controlled operator
\begin{eqnarray}\label{W1}
W^{(1)}_{\chi\psi q K}&=& \prod_{i=1}^n \Big(P_{\chi_iK} \otimes \sigma^{(x)}_{\psi_i} \sigma^{(x)}_{q_i} + (I_{\chi_iK} -P_{\chi_iK}) \otimes  I _{\psi_i q_i}\Big).
\end{eqnarray}
Hereafter the subscript attached to the notation of a subsystem indicates the appropriate qubit of this  subsystem. Thus, subscript $i$  mean  the $i$th  qubit of the appropriate subsystem in (\ref{W1}).

Applying  $W^{(1)}_{\chi\psi q K}$ to $|\Phi_1\rangle$ we obtain

\begin{eqnarray}\label{Phi2}
|\Phi_2\rangle &=& W^{(1)}_{\chi\psi q K} |\Phi_1\rangle = \frac{1}{2^{(n+1)/2}}\left(
\sum_{k=0}^{N-1} q_0 |A\rangle |k\rangle_\chi |k\rangle_\psi |0\rangle_q |0\rangle_K\right. \\\nonumber
&+& \left.\sum_{k=0}^{N-1} q_k |A\rangle |k\rangle_\chi |k\rangle_\psi |0\rangle_q |1\rangle_K)\right) + |g_2\rangle,
\end{eqnarray}
where  the garbage $|g_2\rangle$ collects the terms containing   the states $|j\rangle_q$ with $j>0$, which we don't  need hereafter.

Next, we  prepare and apply the unitary operators  $U(\alpha)$ and $U(\beta)$  in (\ref{obj}) controlled  by the excited state of the one-qubit subsystem $K$:
\begin{eqnarray}\label{W2}
&&W^{(2)}_{\chi\psi K }= |1\rangle_K\, _K\langle 1 |\otimes U_{\chi} (\alpha) U_{\psi} (\beta) +   |0\rangle_K\, _K\langle 0 |\otimes I_{\chi\psi}.
\end{eqnarray}
We can  represent the action of  the operator $U$ on the  vectors $|k\rangle_\chi$ and  $|k\rangle_\psi$ in terms of  its elements  as follows:
\begin{eqnarray}\label{UU}
U_{\chi}(\alpha) |k\rangle_{\chi}  = \sum_{l=0}^{N-1}b_{lk} (\alpha)|l\rangle_{\chi},\;\;
U_{\psi} (\beta) |k\rangle_{\psi}  = \sum_{l=0}^{N-1} b_{lk}(\beta) |l\rangle_\psi.
\end{eqnarray}
Then, applying $W^{(2)}_{\chi\psi K }$ to $|\Phi_2\rangle$ we obtain

\begin{eqnarray}\label{Phi02}
|\Phi_3\rangle &=&W^{(2)}_{\chi\psi K}|\Phi_2\rangle
\\\nonumber
& =&
 \frac{1}{2^{(n+1)/2}}\left(
\sum_{k=0}^{N-1} q_0 |A\rangle |k\rangle_\chi |k\rangle_\psi |0\rangle_q|0\rangle_K\right. \\\nonumber
&+& \left.\sum_{k=0}^{N-1}
\sum_{l_1=0}^{N-1}\sum_{l_2=0}^{N-1} q_k b_{l_1k}(\alpha) b_{l_2k}(\beta) |A\rangle |l_1\rangle_\chi |l_2\rangle_\psi |0\rangle_q |1\rangle_K)\right) +|g_3\rangle.
\end{eqnarray}
Here, the garbage $|g_2\rangle$ from (\ref{Phi2}) is transformed to   $|g_3\rangle$, but we don't describe explicitly this transformation because we are not interested in the particular structure of the garbage.  The same holds for the garbage in other states below.
To  multiply three matrices $U^T_{\chi}(\alpha)$, $A$ and $U_{\psi}(\beta)$
and  eventually calculate the sum
 $\sum_j q_j  \, _\psi\langle j | U^T(\alpha) A U(\beta)|j\rangle_\psi$ in (\ref{obj}), we follow
 Refs.\cite{ZQKW_arxive2023, ZBQKW_2024}. Using projectors (\ref{PP})
and projectors
 \begin{eqnarray}\label{PP2}
P_{\psi_i K} = |1\rangle_{\psi_i} |1\rangle_K \,  _{\psi_i}\langle 1|_K\langle 1|,\;\; i=1,\dots, n,
 \end{eqnarray}
we  introduce the following controlled operators:
 \begin{eqnarray}
 C^{(1)}_{\chi KR} &=& \prod_{i=1}^{n} \Big( P_{\chi_i K} \otimes \sigma^{(x)}_{R_i} +
 (I_{\chi_iK} -P_{\chi_iK} ) \otimes I _{R_i} \Big),\\\nonumber
  C^{(2)}_{\psi K C} &=& \prod_{i=1}^{n} \Big( P_{\psi_iK} \otimes \sigma^{(x)}_{C_i} +
 (I_{\psi_iK} -P_{\psi_iK} ) \otimes I _{C_i} \Big) ,
 \end{eqnarray}
 Here, the operator $C^{(1)}_{\chi K R}$  is required for multiplying $U^T(\alpha)$ and $ A$, the operator
  $C^{(2)}_{\psi K C}$ serves  for multiplying $A$ and  $U(\beta)$,
 Applying the operator $W^{(3)}_{RC\chi\psi  K} = C^{(2)}_{\psi K C} C^{(1)}_{\chi K R}$ to the state
$ |\Phi_3\rangle$  we obtain
\begin{eqnarray}\label{Phi022}
&&|\Phi_4\rangle =W^{(3)}_ {RC\chi\psi  K}   |\Phi_3\rangle
\\\nonumber
&& = \frac{1}{2^{(n+1)/2}}\left(
\sum_{k=0}^{N-1}q_0 a_{00} |0\rangle_R|0\rangle_C |k\rangle_\chi   |k\rangle_\psi |0\rangle_q |0\rangle_K\right. \\\nonumber
&&+ \left.\sum_{k=0}^{N-1}
\sum_{l_1=0}^{N-1}\sum_{l_2=0}^{N-1} q_k b_{l_1k}(\alpha) b_{l_2k}(\beta) a_{l_1l_2}|0\rangle_R|0\rangle_C |\rangle |l_1\rangle_\chi    |l_2\rangle_\psi |0\rangle_q |1\rangle_K)\right)+|g_4\rangle
\end{eqnarray}
where the first part in the rhs collects the terms needed for further calculations  (these terms will be labelled later on by the operator $W^{(5)}_{ R C \chi\psi qB\tilde B}$, see eq.(\ref{W5})) and $|g_4\rangle$ is  the garbage to be removed later.
Now,  according to the multiplication algorithm  (see Appendix in Ref.\cite{ZBQKW_2024}) we introduce  the  operator
\begin{eqnarray}
W^{(4)}_{\chi\psi} = H_{\chi}H_\psi ,
\end{eqnarray}
where  $H_{\chi}$ and   $H_\psi$  are the sets of Hadamard transformations applied to each qubit of the subsystems $\chi$ and $\psi$ respectively. These operators   complete the multiplications  $U^T(\alpha) A$ and $AU(\beta)$ respectively, and  simultaneously   calculate the weighted  trace $\sum_j q_j (U^T(\alpha)AU(\beta))_{jj}$.
Then, applying $W^{(4)}_{\chi \tilde\chi\psi}$ to the state $  |\Phi_4\rangle$, selecting only the needed terms and moving others to the garbage $|g_5\rangle$, we obtain
\begin{eqnarray}\label{Phi023}
|\Phi_5\rangle &=&W^{(4)}_{\chi\tilde\chi\psi}|\Phi_4\rangle
\\\nonumber
& =&
 \frac{1}{2^{(3n+1)/2}}\left(\tilde a_{00}|0\rangle_K + \sum_{l=0}^{N-1} A_l |1\rangle_K\right) |0\rangle_R|0\rangle_C |0\rangle_\chi |0\rangle_\psi |0\rangle_q+|g_5\rangle,
\end{eqnarray}
where,
\begin{eqnarray}\label{tildeAM}
&&\tilde a_{00} =2^{n} q_0  a_{00},\\\label{AM}
&&
A_l (\alpha,\beta)= \sum_{k=0}^{N-1} \sum_{m=0}^{N-1} q_l a_{km}  b_{kl}(\alpha) b_{ml}(\beta) =q_l (U^T(\alpha) M U(\beta))_{ll}.
\end{eqnarray}
Remark that the factor $2^{n}$ in  the expression for  $\tilde a_{00}$  (\ref{tildeAM}) appears  because of the sum over $k$ in  (\ref{Phi022}) which includes $n$ terms.

Now we label and remove the garbage $|g_5\rangle$ from the state (\ref{Phi023})   via the controlled measurement.        To this end we
introduce two  one-qubit ancillae $B$ and $\tilde B$ in the ground state and the controlled operator
\begin{eqnarray}\label{W5}
&&
W^{(5)}_{ R C \chi \psi q B\tilde B} =P\otimes \sigma^{(x)}_B  \sigma^{(x)}_{\tilde B} +
(I_{R C\chi \psi q B\tilde B}-P)\otimes I_{B\tilde B},\\\nonumber
&&
P= |0\rangle_R|0\rangle_C |0\rangle_{\chi}  |0\rangle_\psi
 |0\rangle_{q}\,
 _R\langle 0| _C\langle 0|  _{\chi}\langle 0|  _\psi\langle 0|  _q\langle 0|.
\end{eqnarray}
Then, applying $W^{(5)}_{ R C \chi \psi q B\tilde B}$ to the state $ |\Phi_5\rangle |0\rangle_B|0\rangle_{\tilde B}$ we obtain
\begin{eqnarray}\label{Phi3}
&&
|\Phi_6\rangle = W^{(5)}_{R C\chi\psi qB\tilde B} |\Phi_5\rangle |0\rangle_B|0\rangle_{\tilde B}=
\\\nonumber
&&
  \frac{1}{2^{(3n+1)/2}}\left(\tilde a_{00}|0\rangle_K + \tilde L(\alpha,\beta)|1\rangle_K\right) |0\rangle_R|0\rangle_C |0\rangle_\chi   |0\rangle_\psi   |0\rangle_{q} |1\rangle_B|1\rangle_{\tilde B}\\\nonumber
  &&+|g_5\rangle |0\rangle_B|0\rangle_{\tilde B}, \;\;\tilde L (\alpha,\beta)= \sum_{l=0}^{N-1} A_l(\alpha,\beta)  .
\end{eqnarray}
Now we can remove the garbage by  introducing  the controlled measurement  \cite{FZQW_arxive2025},
\begin{eqnarray}\label{ControlledM}
W^{(6)}_{B\tilde B}= |1\rangle_B \, _B\langle 1|\otimes M_{\tilde B} +  |0\rangle_B \, _B\langle 0|\otimes I_{\tilde B},
\end{eqnarray}
where
$M_{\tilde B}$ is the measurement operator applied to $\tilde B$.
Applying $ W^{(6)}_{B\tilde B}$ to $|\Phi_6\rangle$ we obtain
\begin{eqnarray}\label{Phi4}
&&
|\Phi_7\rangle = W^{(6)}_{B\tilde B}  |\Phi_6\rangle
=
|\Psi_{out}\rangle_{KB} |0\rangle_R |0\rangle_C |0\rangle_{\chi}
   |0\rangle_{\psi}  |0\rangle_{q} ,
   \\\nonumber
   &&
|\Psi_{out}\rangle_{KB}=   G^{-1}\Big(\tilde a_{00}   |0\rangle_K+ \tilde L(\alpha,\beta)
 |1\rangle_K\Big)|1\rangle_B,
\end{eqnarray}
where
$G=\sqrt{\tilde a_{00}^2+ |\tilde L(\alpha,\beta)|^2} $ and we recall that $\tilde a_{00}$ in (\ref{tildeAM}) is real because $a_{00}$ is the real element of the matrix  $A$ and $q_0$ is a positive integer.

{  {\bf Remark.}  To obtain the output state $|\Psi_{out}\rangle_{KB}$ we involve so-called controlled measurement.  Of course, we could obtain this state  just using multiple running of the algorithm each time measuring  the state of the ancilla $\tilde B$ till obtaining the state $|1\rangle_{\tilde B}$ as the required  result of measurement. In this case we again obtain the state $|\Phi_7\rangle$ (\ref{Phi4}) with the output state $|\Psi_{out}\rangle_{KB}$.  However, the probability of such output of the measurement is $O( 2^{-(3n+1)})$ which exponentially decreases with the number of qubits  $n$ characterizing the space of the circuit.  This is a disadvantage of the algorithm that can be completely  overcome referring to the controlled measurement. Although we can not present a particular realization of this operator in terms of well known quantum and/or classical operations, the controlled measurement reflects the deep relation between the  quantum and classical physics.  The measurement of state of the qubit $\tilde B$ will be performed if only the excited state $|1\rangle_B$ exists in the considered  superposition state. Since our superposition state   $|\Phi_6\rangle$ (\ref{Phi3}) includes $|1\rangle_B$  by construction, the measurement must be performed with the predictable result $|1\rangle_{\tilde B}$.  In addition, the controlled measurement is the last element in the following	scheme. We know  the control operations, where  both controlling and controlled subsystems are quantum (CNOT is the simplest representative). The controlled operations with controlling classical subsystem and controlled quantum  subsystem  are also  acknowledged \cite{QFT3}. The proposed concept of controlled measurement represent the last possible type of  control operations in quantum-classical system with the control by the state of a quantum subsystem. All these arguments do not prove realizability of controlled measurement but motivate the research for its  realization  that would not contradict the basic postulates and theorems of quantum mechanics including the no-cloning theorem \cite{KChV}. }

We are aimed at finding the objective function $L$ and normalization $G$.  To this end we  apply the Hadamard transformation $H_B$ to the ancilla $B$ and then apply the Hadamard transformation $H_K$, controlled by the excited state of $ B$, to the qubit $K$, i.e., the controlled operator
\begin{eqnarray}
C_{B K}  = |1\rangle_{B}\, _{B}\langle 1| \otimes H_K +  |0\rangle_{B}\, _{B}\langle 0| \otimes I_K .
\end{eqnarray}
Thus, we have
\begin{eqnarray}\label{tPsiOut}
&&
|\tilde \Psi_{out}\rangle =C_{B K}  H_{B} | \Psi_{out}\rangle_{KB}\\\nonumber
&&=\frac{1}{\sqrt{2}G} \Big(\tilde a_{00}|0\rangle _K+\tilde L(\alpha,\beta) |1\rangle_K \Big) |0\rangle_{B} \\\nonumber
&&
-\frac{1}{2G} \Big( (\tilde a_{00}+\tilde L(\alpha,\beta))|0\rangle_K +(\tilde a_{00}-\tilde L(\alpha,\beta))|1\rangle_K \Big)|1\rangle_{B} .
\end{eqnarray}
Now we measure the both qubits $K$ and $B$ with probabilities $p_{ij}$ for fixing the state $|i\rangle_K |j\rangle_{\tilde B}$, $i,j=0,1$, thus having
\begin{eqnarray}\label{ppp}
&&p_{00}(\alpha,\beta)=\frac{\tilde a_{00}^2}{2 G^2(\alpha,\beta)},\\\nonumber
&&p_{10}(\alpha,\beta)= \frac{|\tilde L(\alpha,\beta)|^2}{2 G^2(\alpha,\beta)} ,\\\nonumber
&&p_{01}(\alpha,\beta)=\frac{G^2(\alpha,\beta)+2 \tilde a_{00} {\mbox{Re}}(\tilde L{ (\alpha,\beta)})}{4G^2(\alpha,\beta)} ,\\\nonumber
&&p_{11}(\alpha,\beta)=\frac{G^2(\alpha,\beta)-2 \tilde a_{00} {\mbox{Re}}(\tilde L{ (\alpha,\beta)})}{4G^2(\alpha,\beta)}.
\end{eqnarray}
From this system we obtain the expressions for  the needed quantities:
\begin{eqnarray}\label{G22}
&&
G^2(\alpha,\beta) = \frac{\tilde a_{00}^2}{2 p_{00}(\alpha,\beta)},\\\label{Lab}
&&
 L(\alpha,\beta)={\mbox{Re}}(\tilde L(\alpha,\beta)) =\frac{\tilde a_{00}}{2p_{00}(\alpha,\beta)}(p_{01}(\alpha,\beta)-p_{11}(\alpha,\beta)).
\end{eqnarray}
{  The depth of the  operator $W^{(5)}_{R C\chi \psi q B\tilde B}$   can be estimated as $O(\log N)$ which is indicated in Fig.\ref{Fig:SVD}. But the depth of the whole algorithm is defined by the operator $W^{(2)}_{\chi\psi K}$  whose depth is also indicated in Fig.\ref{Fig:SVD} and equals $O(Q\log N)$.
This estimation  will be obtained in Sec.\ref{Section:W2W2} after introducing the parameter $Q$.
However, as for the depth of the whole algorithm for calculating the objective function, one can to take into account the probabilistic method implemented  for calculating the objective function using formula  (\ref{Lab}) that includes probabilities $p_{ij}$ ($i,j=0,1$) given in (\ref{ppp}). If the number of runs is $N_r$, then the depth of the algorithm is $ O(N_r Q\log N)$. In turn, $N_r\sim 1/\varepsilon$, where $\varepsilon\ll 1$ is the required precision  for probabilistic calculation of $p_{ij}$.  Therefore, the depth is $O( Q\log(N)/\varepsilon)$. We have to note that, also expression (\ref{Lab}) for the objective function $L$ has $p_{00}$ in the denominator, there is no singularity at $p_{00} \to 0$, because $(p_{01}-p_{11})\to 0$ as well in this case.   However, to provide calculation with required precision $\varepsilon$, we require $p_{00}\gg \varepsilon$. This condition can be replaced by the following one:
\begin{eqnarray}\label{ta00}
\tilde a_{00}^2 = 2^{2n}  q_0^2 a_{00}^2\gg \varepsilon \;\; \Leftrightarrow \;\;
q_0 a_{00}\gg 2^{-n} \sqrt{\varepsilon}.
\end{eqnarray}
 In this case $p_{00} \sim p_{01}\sim p_{11}\gg \varepsilon$, while the probability $p_{10}$ does
not appear in (\ref{Lab}). If the probabilities $p_{00}$, $p_{01}$ and  $p_{11}$ are calculated with the precision
$\varepsilon$, then formula (\ref{Lab}) yields the objective function  $L$ with the precision $\sim \varepsilon$. Then, according to eq.(\ref{obj}), we conclude that the  singular values can be calculated with the precision $\sim \varepsilon/N$.

Now we turn to the case  when the  condition (\ref{ta00}) is destroyed and consider the case $\tilde a_{00}^2\sim \varepsilon$. In particular, $\tilde a$ can be zero.  This is the  case when  we can not use formulae (\ref{ppp}) -- (\ref{Lab}) for calculating the objective function with precision $\varepsilon$ and we have to modify the algorithm.  The simplest way  is  to change the starting point of the algorithm and replace formulae (\ref{Phi0}) and (\ref{Phi1}) with the following one:
\begin{eqnarray}\label{Phi00}
|\Phi_0\rangle=\frac{1}{\sqrt{2}} \Big(   |0\rangle_R |0\rangle_C  |0\rangle_\chi   |0\rangle_\psi |0\rangle_q  |1\rangle_K + |A\rangle |0\rangle_\chi   |0\rangle_\psi |\varphi\rangle_q  |1\rangle_K \Big) = |\Phi_1\rangle.
\end{eqnarray}
After proper  modification of the algorithm we  result in formulae similar to (\ref{ppp}) in which the parameter $\tilde a_{00}$ is  replaced with another parameter that does not depend on the elements of $A$ and $q$. Such formulae allow to calculate the objective function $L$ with any desired precision $\varepsilon$. Further details of the modified algorithm will not be discussed here.
We note that starting equation (\ref{Phi00}) can be used in  case (\ref{ta00}) as well. However, initial state (\ref{Phi0}) is simpler for preparation in comparison with state (\ref{Phi00}) and therefore it is recommended in case (\ref{ta00}). }

 The space required for realization of this algorithm { in both above cases} is $O(\log N)$ qubits.

\noindent

\subsection{Realization of  operator $W^{(2)}_{\chi\psi K}$ for real matrix $A$}
\label{Section:W2W2}
The operator $W^{(2)}_{\chi\psi K}$ in (\ref{W2}) can be conveniently represented as a product of two operators
\begin{eqnarray}\label{W2r}
&&W^{(2)}_{\chi\psi K }=W^{(2)}_{\chi K } W^{(2)}_{\psi K },\\\label{W2r1}
&& W^{(2)}_{\chi K } = |1\rangle_K\, _K\langle 1 |\otimes U_{\chi} (\alpha) +   |0\rangle_K\, _K\langle 0 |\otimes I_{\chi},\\\label{W2r2}
&&W^{(2)}_{\psi K } = |1\rangle_K\, _K\langle 1 |\otimes U_{\psi} (\beta) +   |0\rangle_K\, _K\langle 0 |\otimes I_{\psi}
\end{eqnarray}
as shown in Fig.\ref{Fig:SVD}a.
Notice that two operators $W^{(2)}_{\chi K }$ and $W^{(2)}_{\psi K }$  are completely equivalent to each other  and defer only by the parameters encoded into them.
 Therefore we describe only one of them, say
$W^{(2)}_{\chi K}$.
 To realize the operator $U$ for the real matrix $A$ it is enough to use the one-qubit $y$-rotations
$R_y(\varphi) = \exp(- i \sigma^{(y)} \varphi/2)$ { ($\sigma^{(y)}$ is the Pauli matrix)} and C-nots \cite{WSW}:
\begin{eqnarray}\label{UU2}
&&
U(\alpha) = \prod_{k=1}^Q R_k(\alpha_{(k-1)n +1},\dots,\alpha_{kn}) ,\\\nonumber
&&
R_k(\alpha_{(k-1)n +1},\dots,\alpha_{kn})   = \prod_{m=1}^{n-1} C_{\chi_m\chi_{m+1}}  \prod_{j=1}^{n} R_{y\chi_j}(\alpha_{(k-1)n +j}) ,\\\nonumber
&&C_{\chi_m\chi_{m+1}} = |1\rangle_{\chi_m}\,_{\chi_m}\langle 1| \otimes \sigma^{(x)}_{\chi_{m+1}} +  |0\rangle_{\chi_m}\,_{\chi_m}\langle0 | \otimes I_{\chi_{m+1}} .
\end{eqnarray}
In this formula, $R_k$ represents a single block of transformations encoding $n$ (the number of qubits in the
subsystem  $\chi$) parameters $\alpha_i$, $i=1,\dots, n$, $R_{y\chi_j}$ is the $y$-rotation applied to the $j$th qubit of the
subsystem $\chi$. Involving $Q$ blocks $R_k$, $k=1,\dots, Q$, we
enlarge the number of parameters to $nQ$. We note that this number, in general, { may be} bigger than  the number of free real parameters in the
$N\times N$ unitary transformation, which is $N^2$. Such increase in the number of parameters is caused by
the non-standard parametrization of the unitary transformation $U$ which, in turn, is chosen for two reasons: (i)
simple realization of $U$  in terms of one- and two-qubit operations and (ii) simple realization of  derivatives of the
objective function with respect to these parameters, see eqs.(\ref{derivative}). The depth of the operator $U$ is $O(\log N)$.

Now we turn to realization of the controlled operator $W^{(2)}_{\chi K }$  given in (\ref{W2r1}) . To this end we  substitute $U$ determined in  (\ref{UU2})  into (\ref{W2r1}) and transform it to
\begin{eqnarray}\label{CWW}
&&
W^{(2)}_{\chi K } =\prod_{k=1}^Q \prod_{m=1}^{n-1} C_{K\chi_m\chi_{m+1}}\\\nonumber
&&\times \prod_{j=1}^n
R_{y\chi_j}(\alpha_{(k-1)n+j}/2) C^{(z)}_{K,\chi_j}R_{y\chi_j}(\alpha_{(k-1)n+j}/2) C^{(z)}_{K\chi_j}
\end{eqnarray}
where
\begin{eqnarray}
&&
C_{K\chi_m\chi_{m+1}}= P_{K\chi_m}\otimes \sigma^{(x)}_{\chi_{m+1}}+
( I_{K\chi_m}- P_{K\chi_m})\otimes I_{\chi_{m+1}},\\\nonumber
&&P_{K\chi_m}= |1\rangle_K |1\rangle_{\chi_m}\, _K\langle 1| _{\chi_m}\langle 1|
,
\\\nonumber
&&
C^{(z)}_{K\chi_j}= |0\rangle_K \, _K\langle 0|  \otimes \sigma^{(z)}_{\chi_j}+ |1\rangle_K \, _K\langle 1|
\otimes I_{\chi_j}.
\end{eqnarray}
In (\ref{CWW}), the controlled rotations $R_{y\chi_j}$ are represented by the second product since
\begin{eqnarray}
R_{y\chi_j}(\alpha_{(k-1)n+j}/2) C^{(z)}_{K,\chi_j}R_{y\chi_j}(\alpha_{(k-1)n+j}/2) C^{(z)}_{K\chi_j} =\left\{
\begin{array}{ll}
I_{\chi_j}& {\mbox{for}} \;\; |0\rangle_K\cr
R_{y\chi_j}(\alpha_{(k-1)n+j})& {\mbox{for}} \;\; |1\rangle_K
\end{array}
\right.,
\end{eqnarray}

\begin{figure}[ht]
    \includegraphics[width=0.8\textwidth]{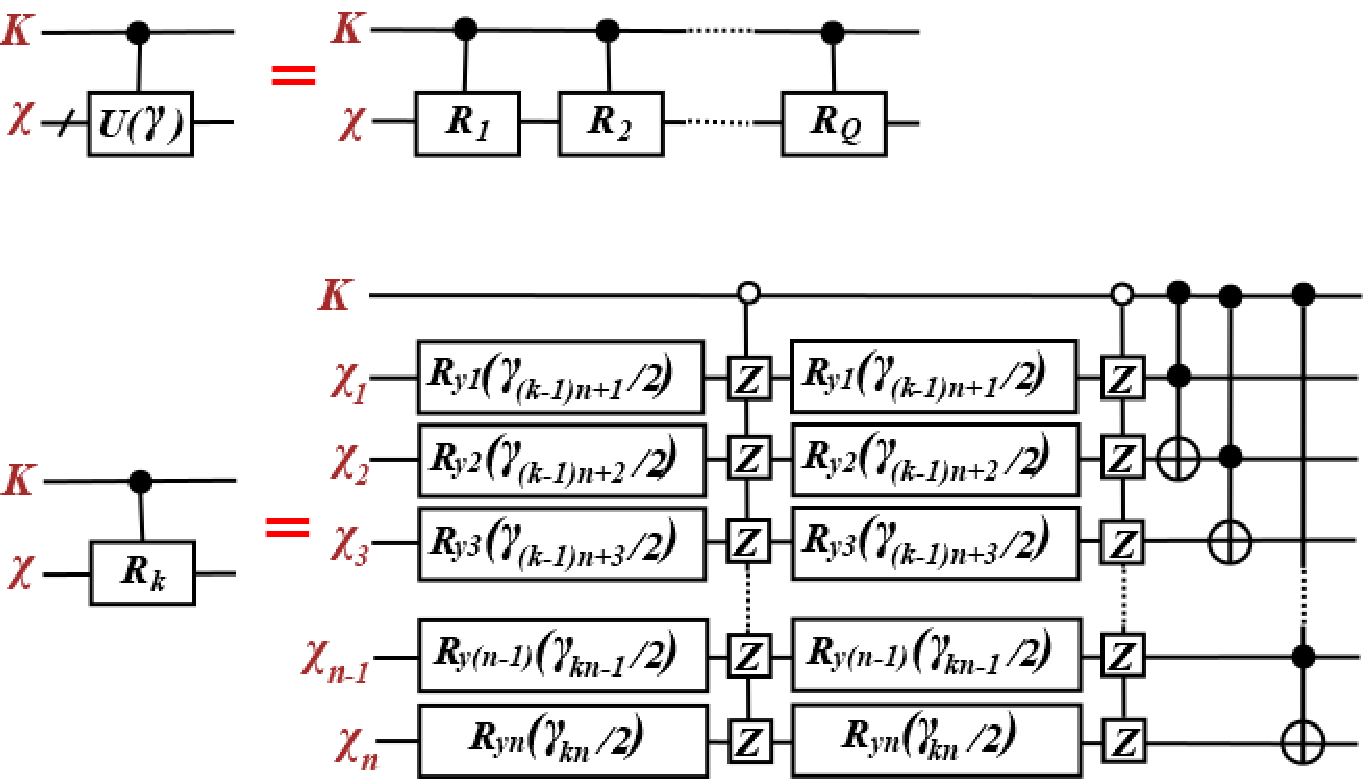}
    \caption{The circuit for the operators  $W^{(2)}_{\chi K}(\alpha) $   (and $W^{(2)}_{\psi K} (\beta)$). Here the set of parameters $\gamma$ is either $\alpha$ (for  $W^{(2)}_{\chi K}(\alpha) $)
     or $\beta$  (for $W^{(2)}_{\psi K} (\beta)$), $Z\equiv \sigma^{(z)}$.}
\label{Fig:Ry}
\end{figure}

The circuit for the operator $W^{(2)}_{\chi K}$ (and also for  $W^{(2)}_{\psi K} $) is shown in Fig.\ref{Fig:Ry}.
{ The depth of each operator $W^{(2)}_{\chi K}$ and $W^{(2)}_{\psi K} $ is $O(Q\log N)$ and therefore the depth of the whole circuit is $O(Q\log N)$, where the parameter $Q$ depends on the required precision of calculating the objective function. The parameter $Q$ is used in the estimation of the depth of the algorithm in the paragraph below eq.(\ref{Lab}).}

{ {\bf Remark.} For the case of complex matrix $A$, the $R_y$-rotation should be replaced with the  operator $
R(\eta,\varphi,\theta) = R_z(\eta) R_y(\varphi) R_z(\theta)$, $R_z=\exp(-i \sigma^{(z)}/2)$, where $\sigma^{(z)}$ is the Pauli matrix.}

\noindent

\subsection{Derivatives of objective function}
\label{Section:der}

The input data for the classical optimization algorithm include not only the value of the objective function at the given  values of the parameters $\alpha$ and $\beta$, but also derivatives of the objective function with respect to these parameters. It  can be shown  \cite{WSW}  that the required derivatives can be obtained calculating the objective function at certain values of the parameters { $\alpha$ and $\beta$} using the algorithm presented in Sec.\ref{Section:general}.

Let $\hat \gamma=\{\hat \gamma_1,\dots,\hat \gamma_{2nQ}\}$ be the set of all parameters $\alpha$ and $\beta$: $\hat\gamma=\{\alpha,\beta\}$.
Since all parameters $\hat\gamma$ are introduced through the $R_y$-rotation, it is simple to calculate  any-order  derivative of $L$ with respect to the parameters $\hat\gamma_j$ \cite{WSW}. For instance, for the first- and second-order derivatives we have
\begin{eqnarray}\label{derivative}
&&
\frac{\partial L(\hat \gamma)}{\partial \hat\gamma_k} = \frac{1}{2}  L(\hat\gamma^{(k)}),\\\nonumber
&&
\frac{\partial^2 L(\hat\gamma)}{\partial\hat\gamma_k \partial\hat\gamma_m} =\frac{1}{4}  L(\hat\gamma^{(k,m)}),\;\;k,m =1,\dots, 2nQ.
\end{eqnarray}
Here $\hat\gamma^{(k)}$ means the set of parameters $\hat\gamma$, in which $\hat\gamma_k$ is replaced with
$\hat\gamma_k+\pi$ and $\hat\gamma^{(k,m)}$ is the set of parameters $\hat\gamma^{(k)}$, in which   $\hat\gamma^{(k)}_m$ is replaced  with
$\hat\gamma^{(k)}_m+\pi$, i.e.,
\begin{eqnarray}
&&
\hat\gamma^{(k)} =\hat \gamma|_{\hat\gamma_k \to \hat\gamma_k+\pi},\\\nonumber
&&\hat\gamma^{(k,m)} =\hat \gamma^{(k)}|_{\hat\gamma_m\to\hat \gamma_m+\pi},\;\;k,m=1,\dots,2nQ.
\end{eqnarray}
In order to probabilistically  find $L(\hat\gamma)$ at fixed values of the parameters, we  have to perform one set of runs of quantum algorithm, the number of runs $N_r$ in this set is determined by the required precision
$\varepsilon$ of $L$: $N_r\sim 1/\varepsilon$.  Similarly, to probabilistically find all first derivatives  $L(\hat\gamma^{(k)})$, $k=1,\dots,2nQ$, at fixed values of  the parameters, we have to perform $2 n Q$ sets of runs (one set for each  derivative). If the optimization algorithm requires also the second  derivatives  $L(\hat\gamma^{(k,m)})$, $k,m=1,\dots,2nQ$, we have to perform
$\frac{2 nQ(2nQ+1)}{2}=2 (nQ)^2+nQ$ sets of runs in addition to the above runs
(we take into account that $L(\hat\gamma^{(k,m)}) = L(\hat\gamma^{(m,k)})$).  The higher order derivatives of the objective function can be treated similarly. In this way we supply the objective function  along with  all necessary  derivatives  of this function to the input of  the classical optimization algorithm which calculates the successive values of the parameters $\hat\gamma$.


\subsection{Hybrid algorithm for SVD: brief discussion}
\label{Section:hybrid}
The variational algorithm for calculating the SVD is a hybrid one. It is described in Refs.\cite{WSW,JKC} in details including examples of realization of the algorithm
 via Paddle Quantum \cite{Paddle1} on the PaddlePaddle Deep Learning Platform \cite{Paddle2,MWW}. The accuracy of the SVD obtained via the variational algorithm  { is estimated by comparing it  with the original matrix.} {  In \cite{WSW}, the authors give detailed analysis of variational quantum SVD (VQSVD) for the randomly generated $8\times 8$ matrix $M$. The quality of VQSVD was characterized by the matrix distance $\|M-M_{SVD}\|_2$ , $M_{SVD}$ is given in (\ref{SVD}), $\|A\|_2=\sqrt{\sum_{i,j} |a_{ij}|^2}$. It was shown that this distance reduces with an increase in the parameter $Q$. 
 Several realizations of the unitary transformation $U$ (see eq. (\ref{UDV})) where given and some  applications of VQSVD (in particula, in Recommendation systems) were proposed.  An alternative VQSVD was proposed in  \cite{JKC} with the principal novelty in encoding the elements of the matrix  $M$ into the quantum state of some system, at that the matrix elements must be ordered according to the certain scheme. The objective function was also different therein.  The quality of VQSVD was characterized by the matrix distance using the Frobenius norm. In comparison with the VQSVD in \cite{WSW}, the modified algorithm demonstrates certain advantages. }

\begin{figure}[ht]
    \includegraphics[width=0.7\textwidth]{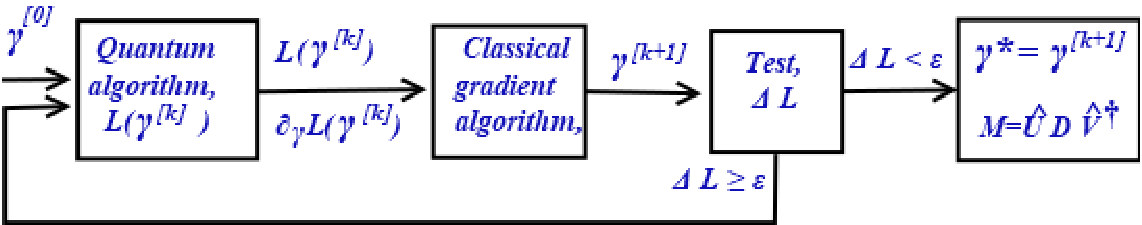}
    \caption{Hybrid algorithm  for calculating SVD.  Matrices $\hat U$, $D$ and $\hat V$ are defined in terms of $U(\alpha^*)$ and $U(\beta^*)$ according to (\ref{UDV}), $\Delta L =| L^{[k+1]}-L^{[k]}|$. For simplicity, we indicate only the function $L$ and first derivatives $\partial_\gamma L$ to be transferred  from the output of the quantum algorithm to the input of the classical algorithm. However, the higher order} derivatives might be required as well.
\label{Fig:QCl}
\end{figure}

The structure of  the hybrid SVD algorithm  is shown in Fig.\ref{Fig:QCl}. We use the superscript $^{[j]}$ to label the $j$th iteration values of the parameters $\hat\gamma$. The algorithm can be briefly described as follow.  For some initial values of the parameters
$\hat \gamma^{[0]}$ we  calculate the values of the objective function $L(\hat\gamma^{[0]})$ and its derivatives with respect to the parameters $\hat\gamma$ using the quantum algorithm. Then we use  the  found values of the objective function and its derivatives    as input for the classical optimization algorithm (for instance, for the gradient maximization algorithm) to find the succeeding iterated  values of the  parameters $\hat\gamma^{[1]}$. Next,  we put them to the input of quantum algorithm, which calculates the objective function and its derivatives for the new values of the parameters $\hat\gamma$ and so on till we reach the required { convergence criterion} $\varepsilon\ll  1$, i.e., till the following condition is satisfied: $\Delta L=|L(\hat\gamma^{[k+1]}) -  L(\hat\gamma^{[k]})|<\varepsilon$. The scheme  for this algorithm is shown in Fig. \ref{Fig:QCl}.
\noindent

\section{Conclusions}
\label{Section:Conclusions}
We present a new version of  the quantum part of the variational SVD algorithm based on the matrix encoding  approach, when the entries of the matrix $M$ are encoded into the probability amplitudes of the superposition state of a quantum system, in our case, subsystems $R$, $C$.  The one-qubit subsystem $K$ is an auxiliary subsystem that controls set of unitary operations. Eventually, the resulting state $|\Psi_{out}\rangle_{KB}$ is the superposition  state  of this auxiliary system $K$ multiplied by the excited state $|1\rangle_B$ of the ancilla $B$.  This state is obtained as a result of controlled measurement of the ancilla $\tilde B$  which removes the problem of small success probability  that unavoidably appears { in the case of ordinary measurement of the ancilla state}  because of the Hadamard transformation used in this algorithm.
{ At the moment, we can not suggest a particular realization of the controlled measurement. However, this operator  means the control of the classical operation (measurement of the qubit $\tilde B$) by the quantum state of another  qubit which is the qubit $B$ in our case, and there is no any  particular  reason to discard possibility of such control.  This control would reflect the deep relation between the classical and quantum phenomena without strong border between them.   The controlled measurement  means that the measurement of $\tilde B$ will be performed if only the superposition state includes the excited qubit $|1\rangle_B$. Otherwise, the quantum system remains unchanged.   Thus, the realization of controlled measurement remains an open problem. It is quite possible that this concept requires some modification to be realizable.}

To measure the value of the objective function we use the  state
 $|\Psi_{out}\rangle_{KB}$  and, after the  Hadamard and controlled Hadamard transformations,  we  find the probabilities of the  states $|i\rangle_{K}|j\rangle_{B} $ and then required objective function $L(\alpha,\beta)$. Since the result is probabilistic we have to run the algorithm many times to obtain the required precision for the objective function. However, this multiple running is necessary part of any probabilistic algorithm. In a similar way we can calculate all derivatives of the objective function required for running the classical optimization algorithm. { The depth of the whole quantum algorithm can be estimated as $O(Q\log(N)/\varepsilon)$, it  linearly depends on the number of runs $N_r$ which, in turn, is inverse proportional to the  precision $\varepsilon$ required for probabilistic measurement of the objective function. The space of the algorithm  is $O(\log N)$.}   We also notice that the different type of the matrix encoding is used in \cite{JKC} yielding certain privileges for that algorithm over the algorithm  in \cite{WSW} .

Although our algorithm deals with square matrices, it can be applied to the rectangular matrices as well because the rectangular matrix can be  written in a square form by adding appropriate number of zero rows or columns.
We also have to note that SVD is also a key for constructing  the inverse or pseudoinverse of the matrix \cite{KP2} because the pseudoinverse matrix
for any given matrix $A$ having SVD,   $A=\sum_{i=1}^rs_i|y_i\rangle\langle x_i|$,  can be written as  $A^+ =  \sum_{i=1}^r\frac{1}{s_i}|x_i\rangle\langle y_i|$.

The fact that matrix-encoding approach is applicable to the variational Quantum SVD algorithm confirms the wide applicability  of this approach  which has already  been used in algorithms for matrix manipulations including addition, multiplication, determinant calculation, inverse matrix calculation and solving systems of linear equations \cite{QZKW_arxive2022,ZQKW_arxive2023,ZBQKW_2024,FZQW_arxive2025}.

{\bf Acknowledgments.} The project is supported by the National Natural Science Foundation of China (Grants No. 12031004, No. 12271474 and No.
61877054). The work was partially funded by a state
task of Russian Fundamental Investigations (State Registration No. 124013000760-0).

\end{document}